# Experimental assessment of critical anthropogenic sediment burial in eelgrass *Zostera marina*

Britta Munkes[1a] & Philipp R Schubert[1a*], Rolf Karez[2], and Thorsten BH Reusch[1]

[1] GEOMAR Helmholtz Center for Ocean Research Kiel, Evolutionary Ecology of Marine Fishes, Düsternbrooker Weg 20, D-24105 Kiel, Germany

[2] State Agency for Agriculture, Environment, and Rural Areas Schleswig-Holstein (LLUR), Hamburger Chaussee 25, D-24220 Flintbek, Germany

[a] B. Munkes and P.R. Schubert each contributed equally to the manuscript (shared first authorship)

* Corresponding author: Philipp R Schubert, pschubert@geomar.de, Düsternbrooker Weg 20, D-24105 Kiel, Germany, Tel.:+49-(0)431-600-4538

## Abstract

Seagrass meadows, one of the world's most important and productive coastal habitats, are threatened by a range of anthropogenic actions. Burial of seagrass plants due to coastal activities is one important anthropogenic pressure leading to decline of local populations. In our study, we assessed the response of eelgrass Zostera marina to sediment burial from physiological, morphological, and population parameters. In a full factorial field experiment, burial level (5-20 cm) and burial duration (4-16 weeks) were manipulated. Negative effects were visible even at the lowest burial level (5 cm) and shortest duration (4 weeks), with increasing effects over time and burial level. Buried seagrasses showed higher shoot mortality, delayed growth and flowering and lower carbohydrate storage. The observed effects will likely have an impact on next year's survival of buried plants. Our results have implications for the management of this important coastal plant.

## Introduction

Seagrass meadows are one of the world's most important and productive coastal habitats (Costanza et al., 1997). They provide a variety of ecosystem services (Hemminga and Duarte, 2000; Larkum et al., 2006) including the stabilization and accretion of sediments in shallow coastal areas (for a review see Fonseca, 1996) as well as some coastal protection by dampening waves and currents (Fonseca and Cahalan, 1992; Fonseca and Koehl, 2006). Seagrass meadows have a positive effect on water transparency, their leaves and rhizomes take up and store nutrients and by that they are reducing eutrophication effects (e.g. McGlathery et al., 2007). Seagrasses provide nursery grounds and shelter for many faunal species, including commercially important fish and shellfish species (for a review see Heck et al., 2003). At the same time, seagrass meadows are threatened worldwide by anthropogenic impacts (eutrophication, overfishing, coastal development, diseases, invasive species, and climate change; e.g. Bockelmann et al., 2013; Munkes, 2005; Orth et al., 2006; Waycott et al., 2009; Williams, 2007).

While eutrophication is considered as the main reason for large-scale losses of seagrass habitats, burial of seagrass plants either through anthropogenic sedimentation or through natural events like storms or mobile sandy bed forms has been identified as an important cause for local die-offs (Cabaço and Santos, 2014; Cabaço et al., 2008; Erftemeijer and Lewis III, 2006; Orth et al., 2006; Short and Wyllie-Echeverria, 1996; Tu Do et al., 2012). Additionally, we can expect to see an increase of sediment mobility due to changes in hydrodynamics as one of the consequences of sea level rise (Dolch and Reise, 2010). This will add to the pressures for seagrasses.

Despite the acknowledged importance, *in situ* experiments testing the effects of increased sedimentation or burial on different seagrass species are rare and short term (for a review of six published studies see Cabaço et al., 2008), and none deals with the most important seagrass of the northern hemisphere, *Zostera marina*, in its core distribution range (Boström et al., 2014).

Along the German Baltic Sea, *Zostera marina* (eelgrass) is the dominant seagrass species (Boström et al., 2014). As an ecosystem engineering species (*sensu* Wright and Jones, 2006) it forms one of the most important coastal habitats in the Baltic Sea. Since the 1950s the areal extent of eelgrass has decreased by at least 51% or 148 km$^2$ along the northern German coast (Schubert et al., 2015), comprising at least one tenth of all known seagrass beds in the area (Boström et al., 2014). In addition to substantial eutrophication accompanied by an increase in turbidity, Baltic eelgrass meadows are experiencing increased sedimentation through beach nourishment, dredging of waterways, discharging of dredged material, and shoreline protection.

For coastal managers and other stakeholders, the height of anthropogenic sediment burial tolerated by eelgrass plants is of particular concern. While sedimentation within existing meadows is in some cases unavoidable as corollary of dredging or beach maintenance, its extent may be manageable.

Additionally, for coastal management and restoration projects, the development of early stress indicators for eelgrass, which hint to existing stressors long before meadows disappear, would be a valuable tool. But despite a variety of existing indicators for seagrass health (Marbà et al., 2013), suitable early stress indicators for eelgrass are still lacking or costly (Macreadie et al., 2014). Storage carbohydrates namely starch and sucrose could have the potential to fill this gap and function as cost-effective indicators for the health status of eelgrass habitats (Govers et al., 2015).

The main aim of this study was to understand the effects of sedimentation level and duration on the seagrass species *Zostera marina*. In contrast to previous studies, we were interested to detect a wide range of eelgrass responses to sediment burial and sedimentation duration. Additionally, our aim was to identify burial threshold levels, above which significant negative effects are likely. Effects on eelgrass plants were measured on three different organizational levels: physiological (carbohydrate content), morphological (shoot biomass, leaf length), and population (shoot density, flowering shoot density, total biomass).

In the present study we investigated the effects of different sediment levels and burial durations within a continuous eelgrass meadow in the western Baltic Sea. We supposed that the relative (% of plant height) rather than absolute (in centimeters) sediment burial level is determining the plants response and survival. Hence, we expected tall eelgrass from the Baltic to be less sensitive to low absolute levels of burial than previously thought (Cabaço et al., 2008).

## Materials and methods

Study area and abiotic environment

The study was conducted in Kiel Bight, a shallow bay of the Baltic Sea with average depths between 16 and 20 m. In Kiel Bight the seagrass species *Zostera marina* is found along most of the coastline in shallow waters between 0.6 and 7.6 m depth (Schubert et al., 2015). It grows between spring (March) and autumn (October) with its peak biomass in late August, early September (Gründel, 1982). In autumn and winter biomass is reduced considerably by storm events, but eelgrass is a perennial species in Kiel Bight. We chose a site called 'Falckenstein' in Kiel Bight for the experiment (Fig. 1, 54°24.388' N, 010°11.575' E). This site features an undisturbed *Zostera marina* meadow with an areal extent of ca. 130 ha within water depths of 0.5 to 5 meters on sandy substratum. The seagrass bed consists mainly of *Zostera marina* (95%), interspersed with some red algae (e.g. *Delessaria sanguinea* and *Ceramium virgatum*) and green algae (e.g. *Ulva* spp.). Also you will find patches of *Mytilus edulis* in the meadow. The eelgrass plants are up to 1.8 m in height during summer, the maximum density during summer is about 500 shoots per m$^2$. It is one of the largest and most consistent eelgrass meadows in this area. At the beginning of the experiment, the water at the study site had a temperature of 9.7 °C. The highest water temperature at the study site was measured on the 30$^{th}$ of July with 23.5 °C. After the 10$^{th}$ of August water temperature slowly decreased down to 17 °C at the termination of the experiment. At water temperatures above 10 °C *Zostera marina* shows normal vegetative growth, so eelgrass plants were not limited in their growth by low water temperatures during our experiment (Reusch et al., 2005). At high water temperatures above 20 °C *Zostera marina* reduces its growth due to heat stress.

The study area is located on the western shore of Kiel Bight, therefore only winds from a northeasterly sector will affect the studied meadow. On the 29th of May (third week of the experiment) strong winds with almost 18 m s$^{-1}$ from east-northeast were measured (GEOMAR data). After this storm most of the sand inside the experimental units was washed out and needed to be refilled during the next week in all units. A second storm from the northeast occurred on the 22nd of July but it was not strong enough to wash out the sand from the experimental units although it had an impact on several of the measured parameters (see 'Results').

Experimental design

We conducted the field experiment in 2014 between spring and late summer, a time of high vegetative growth and maximum biomass of *Zostera marina*. We evaluated the effects of sediment level (sediment height in cm) and experiment duration (time in weeks) on eelgrass by comparing the interactive effects of sediment load and duration on physiological, morphological and population parameters of *Zostera marina*. The complete factorial field experiment was designed with two manipulations:

    **1. Sediment burial** manipulation, with three levels: **control** (= natural height of sediment); '**5 cm burial treatment**' (= addition of 5 cm of sediment) and **'10 cm burial treatment'** (= addition of 10 cm of sediment); and

    **2. Duration of sediment burial**, again with three levels: **4 weeks**; **8 weeks** and **16 weeks**.

Additionally, a further sediment level was included in the experiment as a proof of principle. With this level we tested the hypothesis, that a sediment burial of 20 cm would be lethal for eelgrass plants.

    3. **'20 cm burial treatment'** (= addition of 20 cm of sediment).

This treatment had only one level of duration. All '20 cm burial treatments' lasted 10 weeks, after which the sediment load was removed, followed by a six-week recovery phase. The reason for this recovery phase was the observed high mortality within units of this treatment after 10 weeks.

Each treatment consisted of five replicates summing up to a total of 50 experimental units. The 50 experimental units were arranged inside the meadow (water depth: 2.8 m) in 4 parallel lines to facilitate retrieval and maintenance. The circular units were separated by at least 3 m distance to avoid any potential effects from neighboring units and different treatments were randomly interspersed throughout the study area. Experimental units had a diameter of 40 cm (≙0.126 m$^2$). A random sample of 50 eelgrass plants from the study area taken at the beginning of the experiment indicated an average leaf length of *Zostera marina* of 49.3 cm (SD = 21.3 cm) and an average shoot density of 204 shoots m$^{-2}$ (SD = 40.1 shoots m$^{-2}$). For the experiment, the outer edge of all units was parted off with a knife to cut off all outward rhizome connections. This was done to prevent any connection and possibility for nutrient exchange between plants inside and outside the experimental units.

Each unit, except control units, was contained by a circular plastic lawn edge to keep artificially raised sediment levels constant throughout the experiment. These circular plastic frames were positioned in a sediment depth of ~6 cm. The frames were extended aboveground to the desired height of sediment level (5 cm, 10 cm, and 20 cm). There was no additional shading effect due to the plastic frame. Sediment used for the experiment was purchased from a nearby gravel-pit. We used washed sand for playgrounds in our experiment, as this sand was similar in grain size (coarse sand, 0-1 mm) to sediment at the study site. This ensured that all experimental units were treated with the same quality of sand, free of seeds and any fauna as well as of low organic content to avoid any additional effects due to organic loading. Altogether about 520 kg of sand were used for the experiment. Depending on their treatment, eelgrass plants were buried to 0% (0 cm = controls), ~10% ('5 cm burial treatments'), ~20% ('10 cm burial treatments') and ~40% ('20 cm burial treatments') of the average initial eelgrass height (= 49.3 cm). When scuba divers filled experimental units with sand, care was taken not to break blades and to keep the plants upright during the process.

The experiment started on the 6$^{th}$ of May 2014 and continued for 16 weeks until the 26$^{th}$ of August 2014. Every two weeks eelgrass shoot densities of vegetative and fertile shoots were counted and scuba divers controlled all experimental units. All other parameters (individual shoot and unit biomass, leaf length, carbohydrate content) were sampled at the end of the pre-defined period (after 4, 8, and 16 weeks resp.). For these, all eelgrass shoots including rhizome of one experimental unit were sampled, frozen and stored for later analyses.

Eelgrass parameters

*Shoot density*

Shoot densities of vegetative and fertile shoots were quantified every two weeks of the experiment by counting the total number of shoots inside each experimental unit by scuba divers. Shoot densities in control plots were counted with the help of a frame with the same area as non-control treatments.

*Relative change of shoot number*

Eelgrass plants were considered to be dead when their leaves were black, disintegrating, absent and/or had black rhizomes. The relative change of shoot number (mortality) was calculated as:

$$M (\%) = (d_i - d_f) / d_i \times 100$$

where $d_i$ is the initial shoot density and $d_f$ is the final shoot density (Cabaço et al., 2008).

*Eelgrass biomass*

Individual shoot biomass was determined by drying all eelgrass shoots to a constant weight at 60°C (Kendrick and Lavery, 2001). Total eelgrass biomass per unit was calculated as the sum of all single shoot biomasses of each unit.

*Leaf length*

The average leaf length at the start of the experiment (= 49.3 cm) was estimated by taking a random sample of 50 shoots from within the study area. 12 randomly chosen shoots from this initial sampling were used to calculate the initial maximum leaf length. During the experiment, the length, width and number of all leaves from 12 randomly chosen eelgrass shoots of each experimental unit were measured. From these 12 shoots, the length of the longest leaf was used to calculate the average maximum leaf length per experimental unit.

*Carbohydrate analysis*

Non-structural carbohydrate reserves were quantified after 4, 8, and 16 weeks of the experiment. Sucrose and starch content in eelgrass rhizomes were analyzed according to the method described in Huber and Israel (1982). In short, dried eelgrass material (0.05 g) was ground, the non-structural, soluble, carbohydrates (sugars) from the ground tissues were extracted four times in boiling ethanol (80 °C) and analyzed with the reagent resorcinol (standardized to sucrose). Starch was extracted from the ethanol-insoluble fraction after 12 h at room temperature with the reagent anthrone, again standardized to sucrose (Yemm and Willis, 1954). Finally, absorptions of extracted carbohydrates were measured with a spectrophotometer at an absorbance of 486 (sucrose) and 640 nm (starch) and concentrations were calculated with the standard curve for sucrose.

*Data analysis*

Data of eelgrass vegetative and fertile shoot density were analyzed by two-way repeated measures analysis of variance (ANOVA), with two main effects (factors): sediment level and experiment duration. Data obtained by destructive sampling of experimental units at the end of pre-defined periods (i.e. shoot biomass, total biomass, leaf length, sucrose concentration, starch concentration) were analyzed in two steps: First, we tested the main effects for these parameters for week 4-16 within the sediment levels 'C', '5 cm', and '10 cm'. Data were analyzed by two-way ANOVA, again with the main effects sediment level and experiment duration. Second, we performed an ANOVA for these parameters for week 16 only, but this time including all sediment levels ('C', '5 cm', '10 cm', and '20 cm'). That way, we could keep a completely balanced design for the statistical analyses. Where differences were detected for main effects, we used *a posteriori* multiple comparison test (Tukey-Kramer HSD). All data were transformed when necessary to meet assumptions of normality and homogeneity of variance.

## Results

Shoot density

*Vegetative shoots*
Sediment burial levels and experiment duration both significantly affected eelgrass shoot density, with stronger effects of the factor sediment burial (Table 1). There were also significant interactive effects between both factors. Sediment burials above 5 cm led to mortality (Fig. 2). Units from the 'control treatment' and '5 cm burial treatment' showed a positive net growth rate of eelgrass shoots. Shoot densities in control units increased on average by 37% during the experiment, densities of units with sediment burials of 5 cm increased on average by 30%. Sediment burials of 10 cm or more turned this gain of eelgrass shoots into a loss: burial of 10 cm (20 cm) caused an average mortality of shoots of 5% (87% respectively).

At the beginning of the experiment, control units showed a clear seasonal increase in shoot density until week 10 with an average of 196 shoots m$^{-2}$ and a maximum of 246 shoots m$^{-2}$ (week 10). After week 10, there was a sharp decline in shoot density, which also occurred in the other treatments (Fig. 3-A). After week 12 control shoot densities continuously increased up to 236 shoots m$^{-2}$.

Shoot densities in experimental units of the '5 cm burial treatment' showed a similar pattern to shoot densities in control units with an overall increase during the experiment. Units from '5 cm burial treatment' reached maximum shoot densities (310 shoots m$^{-2}$) also in week 10. Maximum densities were even higher than those of the control units.

Experimental units with a sediment burial of 10 cm showed significantly different shoot densities in comparison to control and 5 cm burial units (Fig. 3-A, Table 1). In these units, there was a continual decrease in average shoot density from 204 shoots m$^{-2}$ to a minimum of 142 shoots m$^{-2}$ in week 12. At the end of the experiment, average shoot density was 163 shoots m$^{-2}$.

The strongest impact was measured in experimental units with a sediment burial of 20 cm. Here, average shoot densities declined continually from an initial value of 248 shoots m$^{-2}$ to minimum densities of 8 shoots m$^{-2}$. This corresponds to a shoot mortality rate of 97% when compared to control units. Even during the 'recovery period', in which the sediment of the '20 cm burial treatment' was removed for the last 6 weeks of the experiment, there was only a slight recovery of shoot densities: at the end of the experiment, maximum densities of 26 shoots m$^{-2}$ were measured.

*Fertile shoots*

The formation of fertile shoots was significantly affected by sediment burial level and experiment duration, whereby sediment burial level had stronger effects on fertile shoot density than experiment duration (Table 1). There were also significant interactive effects between factors. A multiple comparison (post-hoc test) between all treatments showed significant differences between two groups: control units and treatments with 5 cm sediment burial level showed significantly higher densities of fertile shoots than treatments with 10 and 20 cm sediment burial levels.

Fig. 3-B shows the time course of fertile shoots in the controls and different sediment burials. In the beginning of the experiment control units showed a continuous increase in density of fertile shoots. Between week 5 and week 10 maximum densities of 13 fertile shoots m$^{-2}$ were reached. After week 10 a sharp decline in the density of fertile shoots (in line with the decrease of vegetative shoots) occurred. During the last 4 weeks of the experiment control units stagnated at about 3 fertile shoots m$^{-2}$.

Similar to the controls, experimental units with sediment burial levels of 5 cm showed an increase in fertile shoots from the start of the experiment up to week 5. However, '5 cm burial units' only reached maximum densities of 8 fertile shoots m$^{-2}$, which corresponds to a decrease in fertile shoot formation of 39% when compared to control units.

Experimental units with sediment burial of 10 cm showed a lower maximum density of fertile shoots in comparison to control units (5 shoots m$^{-2}$ ≙ 62%). Additionally, a time delay in the formation of fertile shoots occurred. Controls and '5 cm burial units' reached maximum densities of fertile shoots after 5 weeks, while '10 cm burial units' only reached their maximum 2 weeks later (in week 7). Thus, a sediment addition of 10 cm led to a delayed development and a sharp decline in the number of generative shoots and therefore a significantly lower dispersal and recruitment potential. The strongest impact happened in experimental units with sediment burial of 20 cm. In these treatments no fertile shoots formed during the course of the experiment.

Eelgrass biomass

*Total biomass per unit*

Total eelgrass biomass per unit (g DW m$^{-2}$) is shown in Fig. 4-A. We did not find significant differences between sediment burial levels during the first 8 weeks (Table 1). But we found significant effects of sediment burial levels at the end of the experiment. Multiple comparison tests showed that '20 cm burial treatments' and '5 cm burial treatments' were significantly different from all other treatments. Control units and '10 cm burial treatments' had a similar total biomass.

Control units increased in biomass from week 4 to maximum biomass values in week 8 of the experiment, followed by a decrease in biomass towards the end. Experimental units with a sediment burial of 5 cm showed a continual increase in biomass during the entire experiment. However, '5 cm burial treatments' reached their maximum biomass with a delay of 8 weeks. Total biomass of '5 cm burial treatments' equaled 95% of maximum biomass in control units.

The strongest impact was again measured in '20 cm burial treatments'. They showed a decrease in biomass of 80% compared to control plots at the end of the experiment.

*Individual shoot biomass*

Fig. 4-B shows the average biomass (g DW) per eelgrass shoot. We found significant effects of the factor 'sediment burial level' at the end of the experiment. Multiple comparison tests showed that '20 cm burial treatments' were significantly different from all other treatments. In control units, eelgrass shoots reached their maximum biomass in week 8 (0.732 g DW shoot$^{-1}$). Towards the end of the experiment mean biomass of eelgrass shoots slightly decreased due to the storm event and summer recruitment.

Eelgrass shoots from '5 cm burial treatments' obtained their maximal shoot biomass 8 weeks later than eelgrass shoots from control units, reaching comparable shoot biomass levels. Maximum shoot biomass of '10 cm burial treatments' was 32% lower than that of control units. Eelgrass shoots of '20 cm burial treatments' had an average biomass of 0.150 g DW m$^{-1}$, which corresponds to 20% of the average shoot biomass of control units.

Leaf length

Fig. 4-C shows the average of the maximum leaf length of twelve shoots per each experimental unit. Shoots from the control reached their maximum leaf length in week 16 with 86.1 cm.

In '5 cm burial treatments' and '10 cm burial treatments' the maximum leaf length showed only a marginal increase during the first 8 weeks. At the end of the experiment, the longest leaves of both treatments did not reach maximum leaf length of control treatments. The longest leaves of the '20 cm burial treatments' were significantly shorter than all other treatments.

Carbohydrate analysis

*Sucrose concentration*

Experiment duration significantly affected concentration of sucrose in eelgrass rhizomes, whereas sediment burial levels had significant effects only on sucrose content at the end of the experiment in week 16 (Table 1): sucrose content decreased with increasing sediment burial (Fig. 5-A). There was also a significant interactive effect between factors. A multiple comparison between all treatments showed significant differences between the 20 cm sediment burial and all other burial levels. Additionally, we found significant differences between all sampling dates.

Fig. 5-A shows mean concentrations of sucrose in the rhizome of *Zostera marina*. Eelgrass shoots from control units and from '5 cm burial units' showed a similar course, with increasing concentrations of sucrose throughout the experiment and maximum concentrations at the end of the experiment.

Experimental units under a sediment burial of 10 cm, however, showed a delay in carbohydrate storage. From week 4 to week 8, sucrose concentration in the rhizome slightly decreased. 4 weeks later, at the end of the experiment control units and the '5 cm' and '10 cm burial treatments' showed similar concentrations of sucrose. Only experimental units from the '20 cm burial treatment' had significantly reduced sucrose concentrations.

*Starch concentration*

Starch concentration in eelgrass rhizomes was significantly affected by sediment burial level (Table 1). The highest values were found in control units, while starch concentrations of all other treatments were reduced compared to control values. Experiment duration had no significant effect on starch concentration. Also, no significant interactive effects between both factors were measured. A multiple comparison between all treatments showed significant differences between controls and '20 cm burial treatments'.

Fig. 5-B displays starch concentrations in eelgrass rhizomes. Control units showed an increase throughout the experiment. Increase in starch concentration was very low during the 16 weeks of the experiment in the '5 cm burial treatments' and maximum starch concentration at the end of the experiment was 29% lower than final concentrations in control units.

In experimental units of the '10 cm burial treatment' starch concentration remained largely unchanged between the start and the end of the experiment, although showing a marked minimum in week 8. Experimental units with a burial level of 20 cm again showed the lowest concentrations.

## Discussion

Our data demonstrate for the first time the strong effects of anthropogenic or natural sediment loads on *Zostera marina* within its core geographic range in Europe. It is also the first time that sedimentation effects on *Zostera marina* were tested over a time period (16 weeks) sufficiently long to assess seasonal effects. We experimentally tested the impact on different organizational levels of eelgrass, including physiological, morphological, and population parameters. Effects were visible even at low burial levels (5 cm) and short durations (4 weeks), with increasing effects over time and across all levels of organization. The most important findings were increased shoot mortality with increasing sediment load, and delayed leaf growth and flowering as well as lower carbohydrate storage already at low sediment levels of only 10% of eelgrass shoot length.

In their review on burial effects on seagrasses, Cabaço et al. (2008) found that the best predictor of the vulnerability of a seagrass species to burial is its general leaf length. *Zostera marina* is among the seagrass species with the longest leaves in the world (up to 180 cm in our region). The only published burial experiment burial in *Zostera marina* (Mills and Fonseca, 2003) was conducted during 3 weeks in North Carolina (USA) at the southern limit of eelgrass distribution on the east coast of the United States, where eelgrass length averaged only 16 cm at the start of the experiment (as compared to a duration of 16 weeks and an initial height of ~50 cm in our study). Additionally, populations of *Zostera marina* in that study were annual due to high summer temperatures leading to a rapid decline in August and September, the time of the highest biomass in our Baltic Sea meadows. The authors found a substantial reduction of survival and productivity at sediment levels of 25%, and a total die-off at 75% or more of plant height. In our study, we found a substantial reduction of survival at burial levels as low as ~20% of plant height, in line with previous results by Cabaço et al. (2008), who state that species without vertical rhizomes like *Zostera marina* showed very strong negative effects already under low burial levels. At sediment levels of ~40% of plant height shoot mortality averaged 93% after only 12 weeks of burial. Therefore, we even found a higher sensitivity of *Zostera marina* to low levels of sedimentation or burial compared to the earlier study. This higher sensitivity could be due to the longer duration of our experiment (16 vs. 3 weeks).

As we were especially interested to determine threshold values for burial heights of local eelgrass, the question remained as to which extent local long-leaved plants could tolerate anthropogenic burial. Based on our results, we would conclude that negative burial effects on seagrasses depend on the actual leaf length of affected plants rather than general species size. Therefore it is important to consider the actual leaf length of seagrass plants during the sedimentation impact and not to calculate the tolerable sedimentation load based on maximum leaf length of the affected species.

The '5 cm treatment' (~10% burial depth relative to plant height) showed higher maximum 'shoot density' (mean: 310.4 shoots $m^{-2}$), when compared to the control treatment (mean: 246.4 shoots $m^{-2}$). This might be due to an overcompensation of eelgrass plants in response to the increased sedimentation. Many seagrass species commonly show growth responses when threatened by higher sedimentation or burial (Cabaço et al., 2008), e.g. increases in leaf sheath length, vertical internodes length, or branching frequency. Mills and Fonseca (2003) on the other hand did not observe any increases in eelgrass growth parameters in reaction to experimental burial. But they conceded that the short duration of their experiment (24 days) might not have provided sufficient time for growth responses to sedimentation. They recommended a longer study under low conditions of burial to fully understand potential non-lethal responses of eelgrass to sedimentation. With a duration of 16 weeks and sediment levels as low as 10% of the plant height, our experiment complied with these requirements. However, even though maximum shoot density in the '5 cm treatment' was higher compared to controls, we observed a significant delay of 4-8 weeks until eelgrass reached (or surpassed) control levels in most parameters. Furthermore, we found a significant reduction in starch content and flowering capacity, a lower maximum biomass and a lower maximum leaf length of the '5 cm treatment', proving an impact for next years' plants even at low sediment levels. Particularly the reduction in starch content and lower flowering capacity could have strong negative effects on the population level. Govers et al. (2015) found that starch content in the rhizome is a good predictor for winter survival of seagrass plants. Combined with a lower sexual recruitment due to a reduced flowering capacity, the burial could lead to a long-term reduction in eelgrass shoot density.

The '20 cm burial treatment' (~40% of initial plant height) was characterized by very high shoot mortality rates. In one unit, a total die-off occurred and in all other '20 cm treatments' units the plants exhibited extremely low values in all assessed parameters. Although *Z. marina* already showed vulnerability to low levels of sedimentation, it is worth mentioning that in our '20 cm treatment' some plants did survive burial levels that were deadly to most other seagrass species (Cabaço et al., 2008). Only two of the least sensitive species (*Posidonia australis* and *P. sinuosa*) survived burial of 20 cm ≙30-60% of plant height (pers. comm. from J.M. Ruiz in: Cabaço et al., 2008). The recovery phase of 6 weeks conducted for the '20 cm treatments' showed a minor effect on shoot density with an increase in the last 2 weeks of the experiment. This increase is most likely due to lateral ingression of seeds from the surrounding meadow, as indicated by very low maximum leaf lengths, and demonstrates that disturbed areas can be inseminated from intact habitat in the vicinity.

The reason for the observed decline after week 10 in shoot density of all treatments is not entirely clear(Fig. 3-A). Most likely the second storm from the northeast on the 22[nd] of July (week 11) led to the observed decrease, which was also obvious but less pronounced in the density of flowering shoots (Fig. 3-B).

While our experiment showed very strong and negative effects of sedimentation on seagrass health, we are aware of some limitations of our study compared to real life burial of seagrass meadows. Most importantly, we conducted the experiment in a healthy and consistent seagrass meadow. For other, already stressed meadows, this additional stressor of burial could be the one too much and lead to extinction of local meadows. Furthermore, our method of gentle sediment addition (by divers) prevented a sedimentation plume in the water column, which would have added to the observed pressure of sedimentation. Finally, we conducted this experiment during one vegetation period only. It would have been very interesting to see the long-term effects of sedimentation. Particularly because of the observed lower carbohydrate storage, a follow-up in the next year(s) would have been important to assess the survival-rate of affected shoots. For all these reasons, our results are rather conservative in terms of the expected effects of sedimentation on *Zostera marina*.

Energy storage in seagrasses

We measured non-structural carbohydrates (sucrose and starch) in eelgrass rhizomes, because rhizomes are the main reservoir of carbohydrates (Touchette and Burkholder, 2000). Sucrose, in particular, is known to play an important role as short term energy storage for seagrasses, surpassing starch levels 30-fold (Abel and Drew, 1989; Burke et al., 1996), which is why we expected stronger depletion of sucrose compared to starch. Surprisingly however, in our experiment the concentration of sugar changed only slightly in response to different sedimentation treatments when compared to starch. Starch reservoirs are energetically more costly to form and have to be converted back to soluble carbohydrates when needed, while sucrose reserves are immediate available (Touchette and Burkholder, 2000). In shading stress experiments with *Z. marina*, sucrose rather than starch concentrations responded to light limitation and the authors concluded that reservoirs of sugar rather than starch may improve the chance of survival under environmental stresses (Burke et al., 1996; Salo et al., 2015). In contrast to our study, these shading experiments were running for only 3-4 weeks. This time period might be too short to detect changes in starch concentration. In our experiment, clear effects in starch concentration were visible after more than 8 weeks.

In contrast, in the sister species *Zostera noltii* starch was of particular importance as reserve storage and can be used as an indicator for winter survival (Govers et al., 2015; Olivé et al., 2007). This species also showed similar patterns – no effects on soluble sugars but significant effects on starch content of rhizomes among different treatments of sedimentation – as *Z. marina* in the present study (Cabaço and Santos, 2007). In *Z. marina*, starch is probably not mobilized in anoxic roots and thus not transported to the rhizomes, which could explain the low levels we measured there (Smith et al., 1988; Zimmerman and Alberte, 1996). However, even though the actual reasons remain unclear without further research, in sediment treatments with sufficient duration starch seems a more suitable indicator for the physiological state of Zosteraceae than sucrose.

To conclude, our results clearly show the negative consequences of anthropogenic sediment load on eelgrass plants, which is why we strongly advise against dumping sediments in these important and sensitive coastal habitats. Negative effects were already apparent when 5 cm (~10% of shoot length) of sediment were added and were even more pronounced with a sediment addition of 10 cm (~20% of shoot length). As accompanying effects of most anthropogenic sediment altering activities (e.g. increased turbidity and light limitation caused by suspended sediment particles in the water column) were not assessed (and not triggered) in our experiment, impacts of actual sedimentation activities will likely be even more severe. *Zostera marina* has one of the highest light requirements for seagrass species (Erftemeijer and Lewis III, 2006). Therefore, any long-term reduction in light availability will have strong negative effects on *Zostera marina* and if plants are buried in addition, it can potentially be very harmful and lead to extinction of local meadows.

# Figures and tables

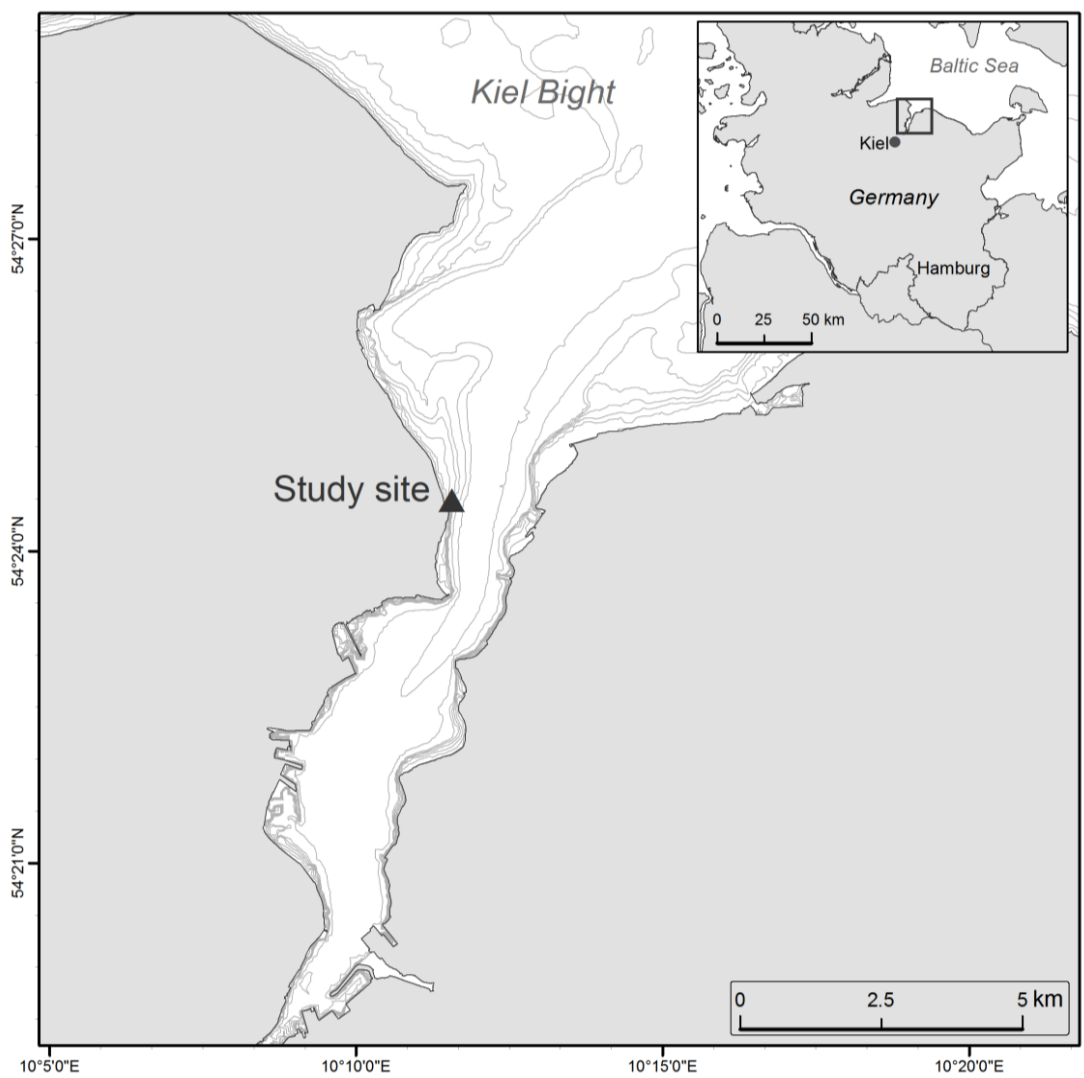

Fig. 1: Map of the study site (black triangle) and surrounding area. Insert: overview of Northern Germany, study area marked with black square.

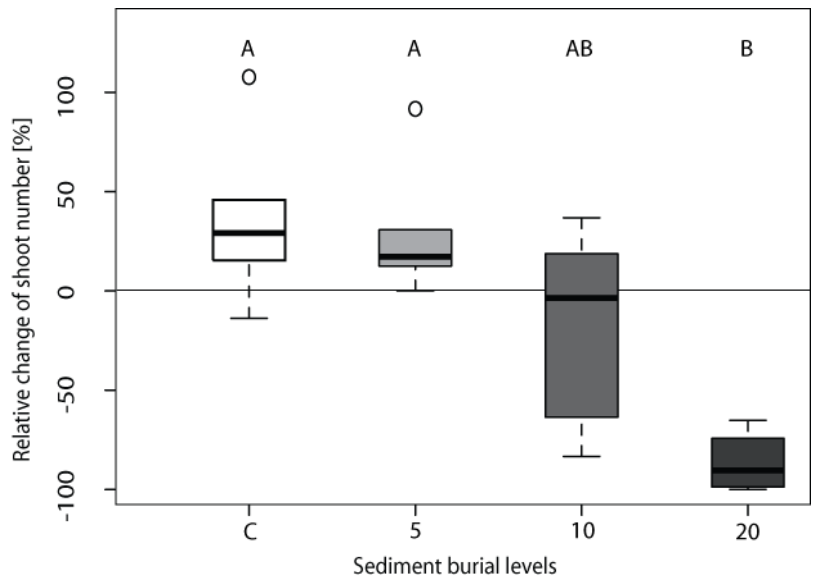

Fig. 2: Box plot of change of shoot number between start and end of experiment by sediment burial level (control, 5, 10, and 20 cm). Centre line: median, box limits: 1$^{st}$ to 3$^{rd}$ quartile, whiskers: range of values. Letters indicate significant results of pairwise Tukey-Kramer HSD *post-hoc* test.

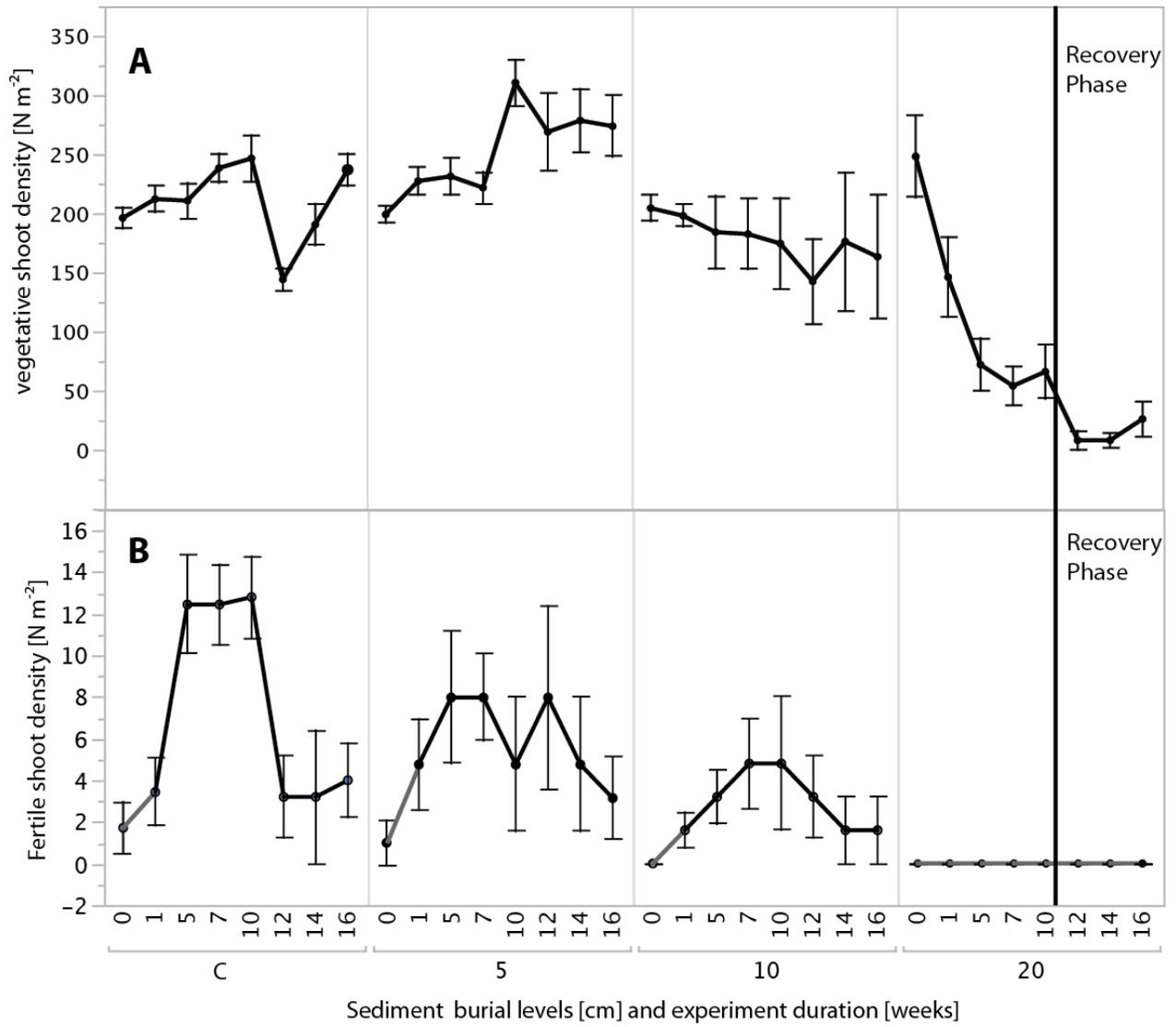

Fig. 3: Means of vegetative (A) and fertile (B) eelgrass shoot density (N m$^{-2}$) by sediment burial level (control, 5, 10, and 20 cm) and experiment duration (0-16 weeks). Bars indicate standard errors (SE).

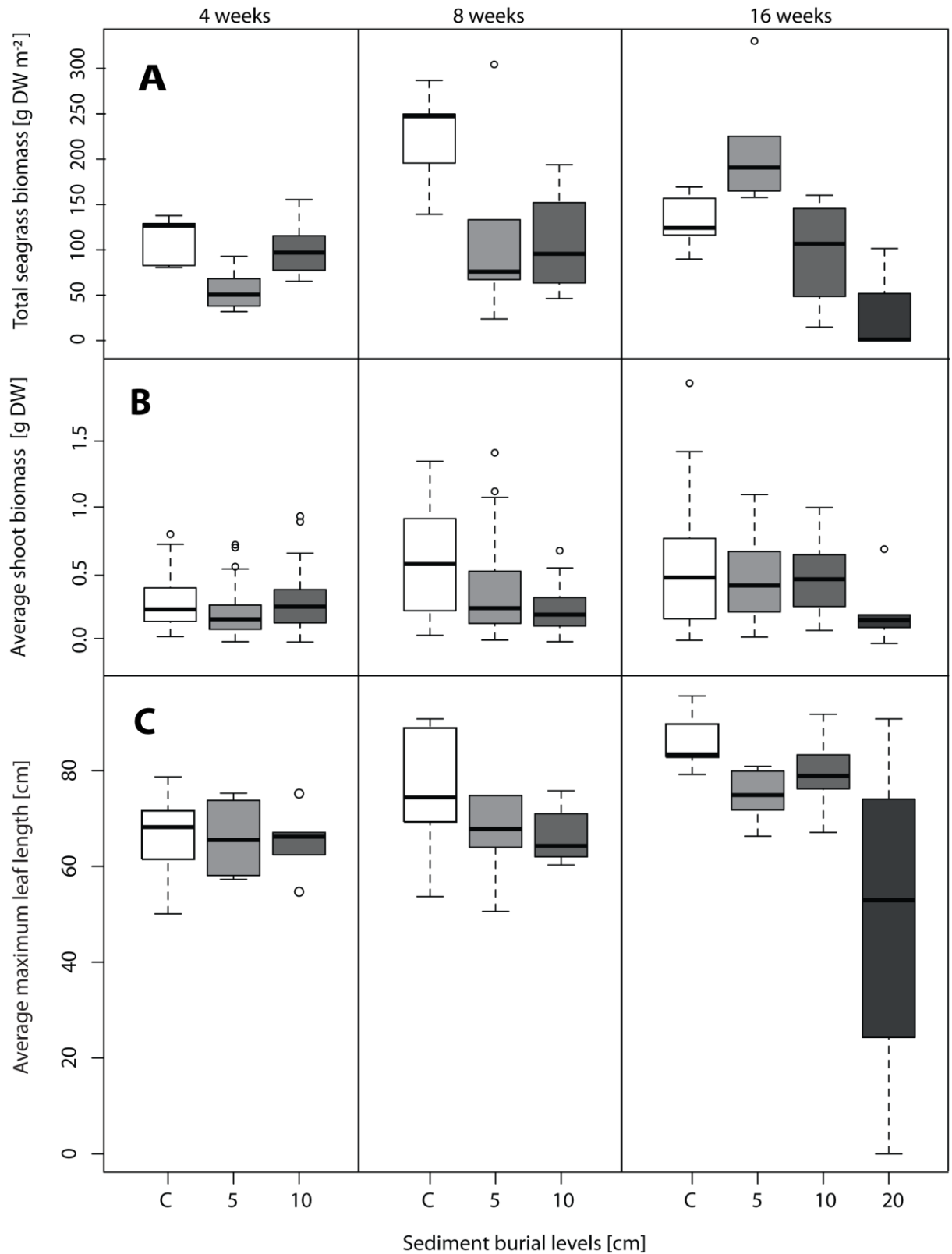

Fig. 4: Boxplots of total biomass (A), average shoot biomass (B), and average maximum leaf length (C) per unit of *Zostera marina* by sediment burial level (control, 5, 10, and 20 cm) and experiment duration (4, 8, and 16 weeks). Centre line: median, box limits: 1[st] to 3[rd] quartile, whiskers: range of values, open dots: outliers.

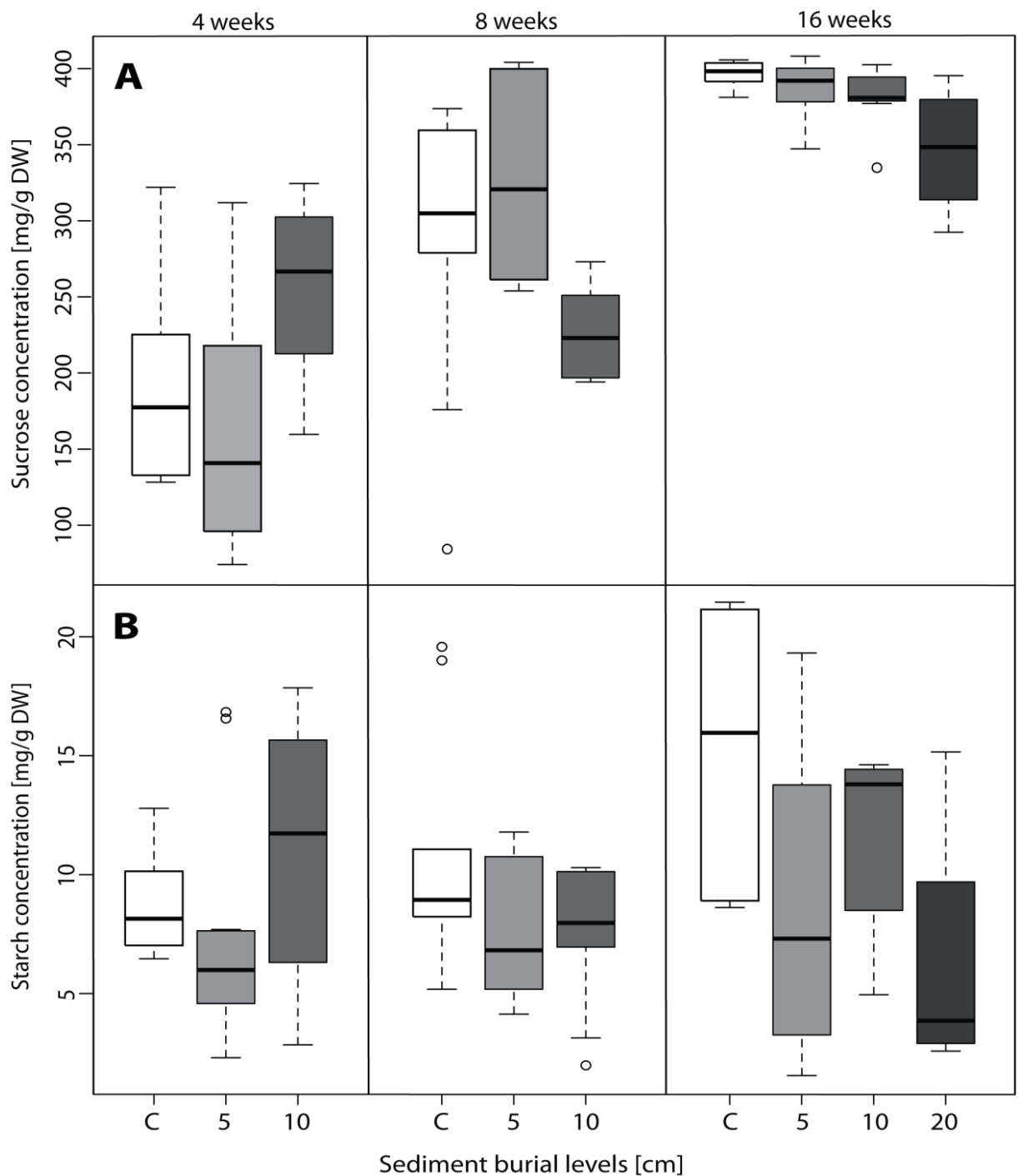

Fig. 5: Boxplots of average sucrose (A) and starch (B) concentrations of *Zostera marina* rhizomes (mg $g^{-1}$ DW) by sediment burial level (control, 5, 10, and 20 cm) and experiment duration (4, 8, and 16 weeks). Centre line: median, box limits: 1st to 3rd quartile, whiskers: range of values, open dots: outliers.

Table 1: Results for two-factorial repeated measures ANOVA testing for significant effects of sediment level and experiment duration on eelgrass vegetative shoot density and fertile shoot density. Results for two-factorial ANOVA testing eelgrass biomass $m^{-2}$, shoot biomass, maximum leaf length, (averaged maximum leaf length of 10 shoots per experimental unit), sucrose and starch concentration in eelgrass rhizomes. p-values below 0.05 were considered significant (*).

| Source of variation | Sediment level | | | | Experiment duration | | | | Sediment level x Experiment duration | | | |
|---|---|---|---|---|---|---|---|---|---|---|---|---|
| | df | SS | F | p | df | SS | F | p | df | SS | F | p |
| **Vegetative shoot density** | 3 | 631832.41 | 62.3681 | **<0.0001*** | 7 | 92619.76 | 3.9182 | **<0.0005*** | 21 | 275742.14 | 3.8884 | **<0.0001*** |
| **Fertile shoot density** | 3 | 1119.72 | 12.1371 | **<0.0001*** | 7 | 844.85 | 3.9247 | **0.0005*** | 21 | 723.12 | 3.9247 | **0.0005*** |
| **Biomass $m^{-2}$ week 4-16** | 2 | 19487.83 | 2.7695 | 0.0769 | 2 | 34185.32 | 4.8582 | **0.0139*** | 4 | 62847.50 | 4.4657 | **0.0052*** |
| **Biomass $m^{-2}$ week 16** | 3 | 10.87 | 3.6968 | **0.0430*** | - | - | - | - | - | - | - | - |
| **Shoot biomass week 4-16** | 2 | 0.12 | 1.4637 | 0.2452 | 2 | 0.48 | 5.8062 | **0.0066*** | 4 | 0.31 | 1.8868 | 0.1346 |
| **Shoot biomass week 16** | 3 | 0.28 | 0.5625 | 0.6485 | - | - | - | - | - | - | - | - |
| **Max. leaf lengths week 4-16** | 2 | 355.58 | 1.3018 | 0.2849 | 2 | 424.86 | 1.5555 | 0.2253 | 4 | 607.88 | 1.1128 | 0.3661 |
| **Max. leaf lengths Week 16** | 3 | 775.42 | 0.8738 | 0.4780 | - | - | - | - | - | - | - | - |
| **Sucrose concentration week 4-16** | 2 | 0.00 | 0.1907 | 0.8268 | 2 | 1.00 | 48.1290 | **<0.0001*** | 4 | 0.20 | 4.9210 | **0.0015*** |
| **Sucrose concentration week 16** | 3 | 0.04 | 7.1743 | **0.0009*** | - | - | - | - | - | - | - | - |
| **Starch concentration week 4-16** | 2 | 1.63 | 4.4139 | **0.0156*** | 2 | 0.24 | 0.6723 | 0.5137 | 4 | 0.55 | 0.7446 | 0.5649 |
| **Starch concentration week 16** | 3 | 3.03 | 4.0179 | **0.0162*** | - | - | - | - | - | - | - | - |

## Acknowledgements

This study was funded by the State Agency for Agriculture, Environment and Rural Areas Schleswig-Holstein (LLUR). We thank one anonymous reviewer for constructive suggestions which improved the manuscript. We thank our scientific divers Florian Huber, Jana Ulrich, Nele Wendländer, and Uwe Schubert for their help in the field experiment. Diana Gill was the biggest help with the analyses in the lab. Susanne Landis helped creating nice graphs alongside everything else, thank you for this.

# References

Abel, K., Drew, E., 1989. Carbon metabolism, in: Larkum, A.W.D., McComb, A.J., Shepherd, S.A. (Eds.), Biology of seagrasses: a treatise on the biology of seagrasses with special reference to the Australian region. Elsevier, New York, pp. 760-796.

Bergmann, N., Winters, G., Rauch, G., Eizaguirre, C., Gu, J., Nelle, P., Fricke, B., Reusch, T.B.H., 2010. Population-specificity of heat stress gene induction in northern and southern eelgrass *Zostera marina* populations under simulated global warming. Molecular Ecology 19, 2870-2883.

Bockelmann, A.-C., Tams, V., Ploog, J., Schubert, P.R., Reusch, T.B.H., 2013. Quantitative PCR reveals strong spatial and temporal variation of the wasting disease pathogen, *Labyrinthula zosterae* in Northern European eelgrass (*Zostera marina*) beds. Plos One 8, e62169.

Boström, C., Baden, S., Bockelmann, A.-C., Dromph, K., Fredriksen, S., Gustafsson, C., Krause-Jensen, D., Möller, T., Nielsen, S.L., Olesen, B., Olsen, J., Pihl, L., Rinde, E., 2014. Distribution, structure and function of Nordic eelgrass (*Zostera marina*) ecosystems: implications for coastal management and conservation. Aquatic Conservation: Marine and Freshwater Ecosystems 24, 410-434.

Burke, M.K., Dennison, W.C., Moore, K.A., 1996. Non-structural carbohydrate reserves of eelgrass *Zostera marina*. Marine Ecology Progress Series 137, 195-201.

Cabaço, S., Santos, R., 2007. Effects of burial and erosion on the seagrass *Zostera noltii*. Journal of Experimental Marine Biology and Ecology 340, 204-212.

Cabaço, S., Santos, R., 2014. Human-induced changes of the seagrass *Cymodocea nodosa* in Ria Formosa lagoon (Southern Portugal) after a decade. Cahiers de Biologie Marine 55, 101-108.

Cabaço, S., Santos, R., Duarte, C.M., 2008. The impact of sediment burial and erosion on seagrasses: a review. Estuarine, Coastal and Shelf Science 79, 354-366.

Costanza, R., dArge, R., deGroot, R., Farber, S., Grasso, M., Hannon, B., Limburg, K., Naeem, S., Oneill, R.V., Paruelo, J., Raskin, R.G., Sutton, P., vandenBelt, M., 1997. The value of the world's ecosystem services and natural capital. Nature 387, 253-260.

Dolch, T., Reise, K., 2010. Long-term displacement of intertidal seagrass and mussel beds by expanding large sandy bedforms in the northern Wadden Sea. Journal of Sea Research 63, 93-101.

Ehlers, A., Worm, B., Reusch, T.B.H., 2008. Importance of genetic diversity in eelgrass *Zostera marina* for its resilience to global warming. Marine Ecology Progress Series 355, 1-7.

Erftemeijer, P.L., Lewis III, R.R.R., 2006. Environmental impacts of dredging on seagrasses: A review. Marine Pollution Bulletin 52, 1553-1572.

Fonseca, M.S., 1996. The role of seagrasses in nearshore sedimentary processes: a review. Estuarine Shores: Hydrological, Geomorphological and Ecological Interactions. Blackwell, Boston, MA, 261-286.

Fonseca, M.S., Cahalan, J.A., 1992. A preliminary evaluation of wave attenuation by four species of seagrass. Estuarine, Coastal and Shelf Science 35, 565-576.

Fonseca, M.S., Koehl, M.A.R., 2006. Flow in seagrass canopies: The influence of patch width. Estuarine Coastal and Shelf Science 67, 1-9.

Govers, L.L., Suykerbuyk, W., Hoppenreijs, J.H.T., Giesen, K., Bouma, T.J., van Katwijk, M.M., 2015. Rhizome starch as indicator for temperate seagrass winter survival. Ecological Indicators 49, 53-60.

Gründel, E., 1982. Ökosystem Seegraswiese: Qualitative und quantitative Untersuchungen über Struktur und Funktion einer *Zostera*-Wiese vor Surendorf (Kieler Bucht, Westliche Ostsee). Christian-Albrechts-Universität Kiel, p. 117.

Heck, K.L., Hays, G., Orth, R.J., 2003. Critical evaluation of the nursery role hypothesis for seagrass meadows. Marine Ecology Progress Series 253, 123-136.

Hemminga, M.A., Duarte, C.M., 2000. Seagrass Ecology. Cambridge University Press, Cambridge, UK.

Huber, S.C., Israel, D.W., 1982. Biochemical basis for partitioning of photosynthetically fixed carbon between starch and sucrose in soybean (*Glycine max* Merr.) leaves. Plant Physiology 69, 691-696.

Kendrick, G.A., Lavery, P.S., 2001. Assessing biomass, assemblage structure and productivity of algal epiphytes on seagrasses. Glogal Seagrass Research Methods. Elsevier, Amesterdam, 199-222.

Larkum, A.W., Orth, R.R.J., Duarte, C.M., 2006. Seagrasses: biology, ecology, and conservation. Springer, Dordrecht, The Netherlands.

Macreadie, P.I., Schliep, M.T., Rasheed, M.A., Chartrand, K.M., Ralph, P.J., 2014. Molecular indicators of chronic seagrass stress: A new era in the management of seagrass ecosystems? Ecological Indicators 38, 279-281.

Marbà, N., Krause-Jensen, D., Alcoverro, T., Birk, S., Pedersen, A., Neto, J., Orfanidis, S., Garmendia, J.M., Muxika, I., Borja, A., Dencheva, K., Duarte, C.M., 2013. Diversity of European seagrass indicators: patterns within and across regions. Hydrobiologia 704, 265-278.

McGlathery, K.J., Sundback, K., Anderson, I.C., 2007. Eutrophication in shallow coastal bays and lagoons: the role of plants in the coastal filter. Marine Ecology Progress Series 348, 1-18.

Mills, K.E., Fonseca, M.S., 2003. Mortality and productivity of eelgrass *Zostera marina* under conditions of experimental burial with two sediment types. Marine Ecology Progress Series 255, 127-134.

Munkes, B., 2005. Eutrophication, phase shift, the delay and the potential return in the Greifswalder Bodden, Baltic Sea. Aquatic Sciences 67, 372-381.

Olivé, I., Brun, F.G., Vergara, J.J., Pérez-Lloréns, J.L., 2007. Effects of light and biomass partitioning on growth, photosynthesis and carbohydrate content of the seagrass *Zostera noltii* Hornem. Journal of Experimental Marine Biology and Ecology 345, 90-100.

Orth, R.J., Carruthers, T.J.B., Dennison, W.C., Duarte, C.M., Fourqurean, J.W., Heck, K.L., Hughes, A.R., Kendrick, G.A., Kenworthy, W.J., Olyarnik, S., Short, F.T., Waycott, M., Williams, S.L., 2006. A global crisis for seagrass ecosystems. Bioscience 56, 987-996.

Reusch, T.B.H., Ehlers, A., Hammerli, A., Worm, B., 2005. Ecosystem recovery after climatic extremes enhanced by genotypic diversity. Proceedings of the National Academy of Sciences of the United States of America 102, 2826-2831.

Salo, T., Reusch, T., Boström, C., 2015. Genotype-specific responses to light stress in eelgrass Zostera marina, a marine foundation plant. Marine Ecology Progress Series 519, 129-140.

Schubert, P.R., Hukriede, W., Karez, R., Reusch, T.B.H., 2015. Mapping and modeling eelgrass *Zostera marina* distribution in the western Baltic Sea. Marine Ecology Progress Series 522, 79-95.

Short, F.T., Wyllie-Echeverria, S., 1996. Natural and human-induced disturbance of seagrasses. Environmental Conservation 23, 17-27.

Smith, R., Pregnall, A., Alberte, R., 1988. Effects of anaerobiosis on root metabolism of *Zostera marina* (eelgrass): implications for survival in reducing sediments. Marine Biology 98, 131-141.

Touchette, B.W., Burkholder, J.M., 2000. Overview of the physiological ecology of carbon metabolism in seagrasses. Journal of Experimental Marine Biology and Ecology 250, 169-205.

Tu Do, V., de Montaudouin, X., Blanchet, H., Lavesque, N., 2012. Seagrass burial by dredged sediments: Benthic community alteration, secondary production loss, biotic index reaction and recovery possibility. Marine Pollution Bulletin 64, 2340-2350.

Waycott, M., Duarte, C.M., Carruthers, T.J.B., Orth, R.J., Dennison, W.C., Olyarnik, S., Calladine, A., Fourqurean, J.W., Heck, K.L., Hughes, A.R., Kendrick, G.A., Kenworthy, W.J., Short, F.T., Williams, S.L., 2009. Accelerating loss of seagrasses across the globe threatens coastal ecosystems. Proceedings of the National Academy of Sciences of the United States of America 106, 12377-12381.

Williams, S.L., 2007. Introduced species in seagrass ecosystems: Status and concerns. Journal of Experimental Marine Biology and Ecology 350, 89-110.

Wright, J.P., Jones, C.G., 2006. The concept of organisms as ecosystem engineers ten years on: Progress, limitations, and challenges. Bioscience 56, 203-209.

Yemm, E.W., Willis, A.J., 1954. The estimation of carbohydrates in plant extracts by anthrone. Biochemical Journal 57, 508-514.

Zimmerman, R.C., Alberte, R.S., 1996. Effect of light/dark transition on carbon translocation in eelgrass *Zostera marina* seedlings. Marine Ecology Progress Series 136, 305-309.